\documentclass[fleqn,usenatbib]{mnras}

\usepackage{newtxtext,newtxmath}

\usepackage[T1]{fontenc}

\DeclareRobustCommand{\VAN}[3]{#2}
\let\VANthebibliography\thebibliography
\def\thebibliography{\DeclareRobustCommand{\VAN}[3]{##3}\VANthebibliography}

\usepackage{graphicx}	
\usepackage{amsmath}	

\usepackage{scalerel,tikz}
\usetikzlibrary{svg.path}
\definecolor{orcidlogocol}{HTML}{A6CE39}
\tikzset{orcidlogo/.pic={
 \fill[orcidlogocol] svg{M256,128c0,70.7-57.3,128-128,128C57.3,256,0,198.7,0,128C0,57.3,57.3,0,128,0C198.7,0,256,57.3,256,128z};
 \fill[white] svg{M86.3,186.2H70.9V79.1h15.4v48.4V186.2z}
 svg{M108.9,79.1h41.6c39.6,0,57,28.3,57,53.6c0,27.5-21.5,53.6-56.8,53.6h-41.8V79.1z M124.3,172.4h24.5c34.9,0,42.9-26.5,42.9-39.7c0-21.5-13.7-39.7-43.7-39.7h-23.7V172.4z}
 svg{M88.7,56.8c0,5.5-4.5,10.1-10.1,10.1c-5.6,0-10.1-4.6-10.1-10.1c0-5.6,4.5-10.1,10.1-10.1C84.2,46.7,88.7,51.3,88.7,56.8z};
}}
\newcommand\orcidicon[1]{\href{https://orcid.org/#1}{\mbox{\scalerel*{
\begin{tikzpicture}[yscale=-1,transform shape]
\pic{orcidlogo};
\end{tikzpicture}
}{|}}}}

\newcommand{\aref}[1]{\hyperref[#1]{Appendix~\ref{#1}}}

\title[Metallicity dependence of the IMF]{The metallicity dependence of the stellar initial mass function}
\author[Tanvir \& Krumholz]{
Tabassum S. Tanvir,$^{1}$\thanks{E-mail:\href{mailto:tabassum.tanvir@anu.edu.au}{tabassum.tanvir@anu.edu.au}}
Mark R.~Krumholz$^{\orcidicon{0000-0003-3893-854X}\,1,2}$\thanks{E-mail: \href{mailto:mark.krumholz@anu.edu.au}{mark.krumholz@anu.edu.au}}
\\
$^{1}$Research School of Astronomy and Astrophysics, Australian National University, Canberra, ACT~2611, Australia\\
$^{2}$Australian Research Council Centre of Excellence in All Sky Astrophysics (ASTRO3D), Canberra, ACT~2611, Australia
}

\date{Accepted XXX. Received YYY; in original form ZZZ}

\pubyear{2015}

\begin{document}
\label{firstpage}
\pagerange{\pageref{firstpage}--\pageref{lastpage}}
\maketitle

\begin{abstract}
Dust is important for star formation because it is the crucial component that couples gas to stellar radiation fields, allowing radiation feedback to influence gas fragmentation and thus the stellar initial mass function (IMF). Variations in dust abundance therefore provide a potential avenue by which variation in galaxy metallicity might affect the IMF. In this paper we present a series of radiation-magnetohydrodynamic simulations in which we vary the metallicity and thus the dust abundance from 1\% of Solar to 3$\times$ Solar, spanning the range from the lowest metallicity dwarfs to the most metal-rich early-type galaxies found in the local Universe. We design the simulations to keep all dimensionless parameters constant so that the interaction between feedback and star-forming environments of varying surface density and metallicity is the only factor capable of breaking the symmetry between the simulations and modifying the IMF, allowing us to cleanly isolate and understand the effects of each environmental parameter. We find that at a fixed surface density more metal-rich clouds tend to form a slightly more bottom-heavy IMF than metal-poor ones, primarily because in metal-poor gas radiation feedback is able to propagate further, heating somewhat larger volumes of gas. However, shifts in IMF with metallicity at a fixed surface density are much smaller than shifts with surface density at fixed metallicity; metallicity-induced IMF variations are too small to explain the variations in mass-to-light ratio reported in galaxies of different mass and metallicity. We, therefore, conclude that metallicity variations are much less important than variations in surface density in driving changes in the IMF and that the latter rather than the former are most likely responsible for the IMF variations found in early-type galaxies.

\end{abstract}




\defcitealias{2022MNRAS.516.5712T}{Paper~I}

\section{Introduction}
The stellar initial mass function (IMF) is the mass distribution of stars at the point of their birth. The IMF is one of the most important distributions in astrophysics because it at least partly determines everything from chemical evolution to the strength of feedback during galaxy formation. The IMF has been found to be nearly universal in the Milky Way and its closest neighbours, which are the only locations where measurements of the IMF by direct star counting are possible \citep[and references therein]{2014prpl.conf...53O}. A number of hypotheses have been proposed to explain this lack of variation, mostly focusing on the universality of turbulence \citep{1997MNRAS.288..145P,2002ApJ...576..870P,2008ApJ...684..395H,Hennebelle09a,2012MNRAS.423.2016H,Hopkins13a,Nam21a} and the stabilising effects of stellar radiation feedback \citep[e.g.,][]{Bate09a, Bate12a, Krumholz11e, Krumholz12b,2016MNRAS.458..673G, 2018MNRAS.476..771C} or the isothermal-adiabatic transition \citep[e.g.,][]{2018A&A...611A..89L,2019ApJ...883..140H}. One important outcome of these studies is that, while turbulence alone can explain why the high mass portion of the IMF always has the same slope, some additional physical process is likely required to explain the existence of an IMF peak at a particular mass \citep[e.g.,][]{2014PhR...539...49K, Krumholz16c, 2016MNRAS.458..673G, Guszejnov20a}. This finding is particularly significant because tentative evidence has started to emerge that in the extreme star forming environments found in early-type galaxies (ETGs), the IMF peak shifts slightly towards lower masses than the IMF peak found in the local Universe. There are multiple lines of evidence for such a shift derived from different measurement techniques -- spectroscopic \citep[e.g.,][]{2010Natur.468..940V, Spiniello12a, La-Barbera13a, Conroy17a}, dynamical \citep[e.g.,][]{2012Natur.484..485C, Newman17a, Oldham18a}, and gravitational lensing \citep{Treu10a, Spiniello15a} -- all pointing towards at least qualitatively consistent conclusions; see \citet{Smith20a} for a comprehensive review. There is also much more tentative evidence for the possibility that the IMF might be more bottom-light in ultra-faint dwarf galaxies (e.g., \citealt{Geha13a, Gennaro18a}; however, see \citealt{El-Badry17a} for a contrary perspective).

These observations raise the theoretical question of which physical processes might be responsible for inducing the shift in the IMF peak in ETGs, and at least potentially in dwarfs as well. In \citet[hereafter \citetalias{2022MNRAS.516.5712T}]{2022MNRAS.516.5712T}, we performed  a series of carefully controlled radiation magnetohydrodynamic (RMHD) simulations to study the role of the environment and its interaction with stellar feedback mechanisms in setting the IMF peak. We control these experiments by keeping all the dimensionless parameters constant (e.g., virial parameter, Mach number, Alfv\'en Mach number) while altering only one other parameter so that it is easier to deduce the role of that parameter along with the feedback mechanism. In \citetalias{2022MNRAS.516.5712T} we explored the role of surface density and showed that with increasing surface density the IMF peak shifts towards lower masses than the ones found in the Milky Way, but by less than would be expected purely from a shift in the mean Jeans mass with surface density because of the enhanced effectiveness of stellar radiation feedback in denser environments. The resulting shift in the IMF peak plausibly matches what is required to explain the mass-to-light ratios measured in ETGs.

However, in \citetalias{2022MNRAS.516.5712T}, we did not alter another potentially important parameter that differs between ETGs and other galaxies: the metallicity, and therefore dust properties. Dust matters since it is the component of the ISM that couples gas to stellar radiation. Some previous authors have conducted numerical simulations to understand how changing metallicity, and therefore changing dust abundance, might alter the IMF. For example, \cite{Bate19a} carries out radiation hydrodynamical simulations to study the metallicity-dependence of the properties of stellar populations formed at metallicities $Z = 0.01 - 3$ $\rm Z_{\odot}$, and finds that the stellar mass distribution is largely insensitive to metallicity. This result is in qualitative agreement with a number of similar, earlier numerical studies \citep{Myers11a, 2014MNRAS.442..285B,Bate15a}. More recently, however, \citet{2022MNRAS.515.4929G} found that the characteristic mass of a forming stellar population does vary with both metallicity and the strength of the interstellar radiation field (ISRF). They find that with a typical Galactic ISRF, lower metallicities produce a higher characteristic mass. \cite{2021MNRAS.508.4175C} also studied the IMF in metal-poor environments and found that the number of low-mass stars increases with metallicity and that the mass function is top-heavy in low-metallicity environments. \citet{2022MNRAS.509.1959S} use analytic models to survey a wide range of parameter space, and  find that the characteristic stellar mass is comparatively high at metallicities low enough that metal line cooling dominates, but begins to decrease with metallicity once the metallicity is high enough ($Z\gtrsim 0.01\mathrm{Z}_\odot$) for dust-gas coupling to dominate gas thermodynamics \citet{2023MNRAS.519..688B} conducts simulations taking into account how the combined effects of increasing cosmic microwave background intensity at high redshift and variation in metallicity impacts the IMF. He finds that for the CMB intensity at redshift $z = 5$, increasing metallicity increases the characteristic mass of stars. 

While all these authors have studied the effects of varying metallicity, none of the previous simulations (as opposed to analytic models) have done so for the very high surface-density environments that likely characterised star formation in ETGs. Moreover, the approach in most of these papers was to consider a wide range of environmental changes -- for example not just metallicity, but also CMB intensity, ISRF intensity, total star-forming cloud mass, etc. While this has the advantage of being maximally realistic since in fact all of these parameters do vary with redshift and galactic environment, it has the disadvantage of making the results extremely difficult to interpret. Since these simulations vary many parameters at once, it is not clear which of the many possible knobs that are being turned is responsible for any particular aspect of the simulation outcomes, nor is it easy to identify by what mechanism any particular knob operates. 

In this paper we take a different approach, following on from that in \citetalias{2022MNRAS.516.5712T}: we present a series of RMHD simulations where we explore the role of metallicity variation in setting the IMF by keeping the dimensionless parameters that describe our star-forming clouds constant except for parameters that we vary systematically and one at a time so that we can create a clean experiment that isolates the effects of environment and its interaction with stellar feedback. A critical aspect of our simulation approach is that our experiments are engineered so that, except for the one change that we make, our simulations are all simply re-scaled versions of one another, such that if we were to turn off stellar feedback and complex thermodynamics, and simply adopt an isothermal equation of state, the results of all simulations would be identical. This allows us to solve the problem of interpretability since we can then unambiguously connect effects and causes.

We describe the numerical method and initial conditions we use to achieve these effects in \autoref{sec:methods}. In \autoref{sec:results} we discuss the results from our simulations and their implications for the physics behind the IMF. We summarise our conclusions in \autoref{Conclusion}.

\section{Numerical Methods and Initial Condition}
\label{sec:methods}

\subsection{Numerical Methods}

The numerical methods we employ in this study are identical to those used in \citetalias{2022MNRAS.516.5712T}, and we refer interested readers to that paper for more detailed information. Here we provide a brief summary. We carry out our simulations using the {\sc orion2} adaptive mesh refinement code \citep{Li21b}. The code uses the approach of \cite{2012ApJ...745..139L} to solve the equations of ideal magnetohydrodynamics with self-gravity \citep{1998ApJ...495..821T,1999ASSL..240..131K} and radiation transfer \citep{2007ApJ...656..959K} in the two-temperature, mixed-frame, grey, flux-limited diffusion approximation. 

The code includes sink particles \citep{2004ApJ...611..399K} to replace regions where protostars are forming and that are collapsing beyond our ability to resolve. Each sink particle runs a one-zone protostellar evolution model as described in \citet{2009ApJ...703..131O}, which provides the instantaneous properties of the star that it represents, such as radius, luminosity, and polytropic index; these depend on the accretion history determined self-consistently from the simulations. The luminosity of each sink particle is then used as a source term in the radiative transfer equations. We also take into account the feedback caused by protostellar outflows through momentum sources around each sink particle. The outflow model we use is described in \citet{2011ApJ...740..107C}. In this model, whenever mass is accreted onto a sink particle, a fraction $f_w$ of it is ejected back into the simulation in the form of an outflow. This outflow material is launched with a speed of $v_w$. In our simulation, we adopt the same wind model parameters used in \citet{2012ApJ...747...22H} and \citet{2018MNRAS.476..771C}, with $f_w=0.3$ and $v_w=\min(v_{\rm kep},60\;\mathrm{km/s})$, where $v_{\rm kep}$ is the Keplerian speed at the protostar's surface (also determined self-consistently from the accretion history via the protostellar evolution model). These parameters are based on observations of the momentum budget of protostellar outflows \citep{2000prpl.conf..867R}.

A critical component of the simulations is the opacity of the dusty gas, which is responsible for coupling the gas flow to the radiation field generated by the stars and by the thermal radiation from the dusty gas itself. As in \citetalias{2022MNRAS.516.5712T}, we take the Rosseland and Planck mean opacity of the dusty gas as a function of density and temperature from the tabulated results of \citet{Semenov03a}. In order to study the effects of varying the metallicity, we scale these tabulated opacities by the metallicity relative to Solar, i.e., for runs with metallicity $Z = 0.1 Z_\odot$ we take the opacities to be 10\% of \citeauthor{Semenov03a}'s tabulated values. We should therefore understand the metallicity we quote when describing our simulations as the dust metallicity; the gas-phase metallicity is unimportant as long as the density is high enough for dust and gas to be thermally coupled by collisions since in this case dust heating and cooling completely dominates the gas thermodynamics. 

There is a limitation to the radiation transfer method: we assume the dust and gas temperatures are equal. This is generally a good assumption for gas and dust temperature at densities above $\rm 10^{4}-10^{5} cm^{-3}$ \citep[e.g.,][]{2001ApJ...557..736G}. However, in low-density regions, dust-gas collisions may occur too infrequently to allow efficient gas-dust coupling, allowing gas to be either hotter or cooler than the dust. This is however unlikely to be a factor in determining the IMF since in regions where gas is collapsing and fragmenting, the density is high enough for gas and dust to be well coupled. Studies that have treated gas and dust temperatures separately \citep{Bate15a} find that its effect on fragmentation is minimal \citep{Bate19a}. 

\subsection{Initial and boundary conditions}

We construct our initial conditions following \citetalias{2022MNRAS.516.5712T}, and we refer readers to that paper for full details. Our initial conditions are designed to produce a carefully controlled experiment, whereby we hold all simulation parameters fixed except for one, which we vary systematically in order to isolate the effects of that parameter. In this study, we consider two different series of runs. The ``medium'' density case (M runs hereafter) consists of a periodic box\footnote{While all MHD quantities use periodic boundaries, the radiation field uses Marshak boundaries, with a inward radiation flux corresponding to that of an isotropic blackbody radiation field with a radiation temperature of 10 K; see \citetalias{2022MNRAS.516.5712T} for full details.} containing a mass $M_{\rm box} = 1000$ M$_\odot$ with surface density $\Sigma = 1$ g cm$^{-2}$ (corresponding to a mean density $\rho_0=7.0\times 10^{-19}$ g cm$^{-3}$, box length $L = 0.46$ pc), an initial temperature $T_0 = 10$ K, an initial gas velocity dispersion $\sigma = 2.4$ km s$^{-1}$, and an initially-uniform magnetic field of strength $B_0 = 0.73$ mG; for this combination of parameters, the box free-fall time $t_\mathrm{ff} = 160$ kyr, the crossing time $t_\mathrm{cross} = 0.4$ Myr, the virial ratio $\alpha_\mathrm{vir} = 1$, the Mach number $\mathcal{M} = 12.6$, and the plasma $\beta=0.012$. The ``low'' density series (L runs hereafter) are a rescaled version of the M runs with a surface density that is lower by a factor of $f=1/2$, and a box length, volume density, and magnetic field strength that are multiplied by factors of $f^{-1}$, $f^2$, and $f$ relative to the M series, respectively; the gas velocity dispersion and temperature are unchanged. These transformations have the property that they leave $\alpha_\mathrm{vir}$, $\mathcal{M}$, and $\beta$ unchanged between the L and M series, so the \textit{only} dimensionless number that varies between the L and M series is the optical depth of the gas.

In addition to varying surface density and thus the optical depth at fixed metallicity, as in  \citetalias{2022MNRAS.516.5712T}, in this study we also vary the metallicity independently of the surface density. The metallicity range we explore is from 1\% of Solar to 3$\times$ Solar; this covers the range from the lowest metallicity dwarf galaxies in the local Universe \citep{2013PASP..125..600M} to the most metal-rich early-types \citep{2021arXiv211011985G}. We carry out runs at $Z/Z_\odot = 1\%$, $10\%$, $1$, and $3$ for both the L and M cases; we refer to these runs as L1\%, L10\%, L1$\times$, and L3$\times$, and similarly for the M series, and we summarise their full properties in \autoref{table:simparam}. Note that the L$1\times$ and M$1\times$ runs are identical to the L1 and M1 runs in \citetalias{2022MNRAS.516.5712T}.

Metallicity variation will change how the gas interacts with stellar radiation feedback since metals (in the form of dust grains) are what couple the radiation to the gas. They will also change how the gas interacts with protostellar outflow feedback, since outflows shock the gas, and the rate at which the gas is then able to cool back down to its equilibrium temperature depends on the box optical depth. Since the dimensionless parameters are the only ones that describe the system in the absence of stellar feedback (radiation and protostellar outflow), any differences we find between the simulations can only be a result of the interaction of stellar feedback with the differing surface density and metallicity, and thus the differing optical depth. By comparing to the Solar metallicity runs reported in \citetalias{2022MNRAS.516.5712T}, we can further separate the effects of metallicity and surface density.

\begin{table*}
\centering
\begin{tabular}{ccccccccccccccccc}
\hline \hline
Name & $M_{\rm box}$ (M$_{\odot}$) & $L$ (pc) & $\rho_0$ (g cm$^{-3}$) & $B_0$ (mG) &
$t_{\rm ff}$ (kyr) & $\Delta x$ (AU) & $\rm t_{\rm cross}$ (Myr) & $\Sigma$ (g cm$^{-2}$) & $Z/Z_\odot$ & $\tau_{100K}$ \\
\hline \hline
L1\% & 2000 & 0.92 & $\rm 1.74\times 10^{-19}$& 0.36 & 160& 46& 0.4 & 0.5 & 0.01 & $5.38 \times 10^ {-3}$ \\

L10\% & 2000 & 0.92 & $\rm 1.74\times 10^{-19}$& 0.36 & 160& 46& 0.4 & 0.5 & 0.1 & 0.054\\

L$1\times$ & 2000 & 0.92 & $\rm 1.74\times 10^{-19}$& 0.36 & 160& 46& 0.4 & 0.5 & 1 & 0.54\\

L$3\times$ & 2000 & 0.92 & $\rm 1.74\times 10^{-19}$& 0.36 & 160 & 46& 0.4 & 0.5 & 3 & 1.6\\

M1\% & 1000 & 0.46 & $\rm 6.96\times 10^{-19}$ & 0.73 & 80 & 23& 0.2 & 1 & 0.01 & 0.0108 \\

M10\% & 1000 & 0.46 & $\rm 6.96\times 10^{-19}$ & 0.73 & 80& 23& 0.2 & 1 & 0.1 & 0.108 \\

M$1\times$ & 1000 & 0.46 & $\rm 6.96\times 10^{-19}$ & 0.73 & 80 & 23& 0.2 & 1 & 1 & 1.08 \\

M$3\times$ & 1000 & 0.46 & $\rm 6.96\times 10^{-19}$ &0.73 & 80 & 23& 0.2 & 1 & 3 & 3.2 \\

\hline \hline
\end{tabular}
\caption{Simulation parameters, from left to right: run name, mass in the computational box, size of the computational box, mean density in the computational box, initial magnetic field strength, mean-density free-fall time, cell size at the finest AMR level, turbulent crossing time, surface density, metallicity, and optical depth computed using the Planck-mean opacity evaluated a temperature $T=100$ K. All simulations use the same initial velocity dispersion $\sigma_0 = 2.4$ km s$^{-1}$ and temperature $T_0 = 10$ K.}
\label{table:simparam}
\end{table*}

\subsection{Resolution, refinement, and sink particles}

In order to ensure that we can compare our current runs to those carried out in \citetalias{2022MNRAS.516.5712T}, we use identical resolution, refinement, and sink particle creation criteria, and we refer to that paper for a detailed description, which we only summarise briefly here. The AMR hierarchy in these simulations is set on a $512^{3}$ base grid which we denote $\mathcal{L} = 0$. The simulation takes place in two stages; during the first, ``driving'' phase, with lasts for two crossing times, we disable self-gravity and radiation transport, and drive the turbulence to a steady velocity dispersion, providing time for the turbulence to achieve a statistically steady state. During the driving phase we disable AMR, so no higher level exists. After the driving phase we start the ``collapse'' phase, during which we disable driving and re-enable self-gravity and radiation. Once we turn on gravity the grid is allowed to adaptively refine to a maximum level $\rm \mathcal{L}_{max} = 2$, and we refine in any cell where the Jeans number
\begin{equation}
    J = \sqrt{\frac{G\rho \Delta x^{2}}{\pi c_{s}^{2}}}
\end{equation}
rises above $J=1/8$; here $\rho$ is the gas density, $\Delta x$ is a cell size, and $c_s$ is the gas isothermal sound speed. We report the size of the cells on the finest level in \autoref{table:simparam}. 

Sink particle formation is triggered in any zone on the finest AMR level where the gas is dense enough to reach a local Jeans number $J>1/4$. Once a sink particle is formed it interacts with the gas via gravity, accretion, and stellar feedback only.

\section{Results}
\label{sec:results}

We present the results of our simulations here. First, we give an overview of the simulations in \autoref{ssec:overview}. Next, we discuss the mass distribution of the sink particles and identify how different metallicity at a fixed surface density impacts the IMF in \autoref{ssec:massdist}. In \autoref{ssec:radiation} and \autoref{ssec:outflow} we interpret these results in terms of the effects of radiation and protostellar outflow feedback, and the interaction of these two feedback mechanisms with gas of varying metallicity.

\subsection{Overview of simulations}
\label{ssec:overview}
\begin{figure*}
	\includegraphics[width=\textwidth]{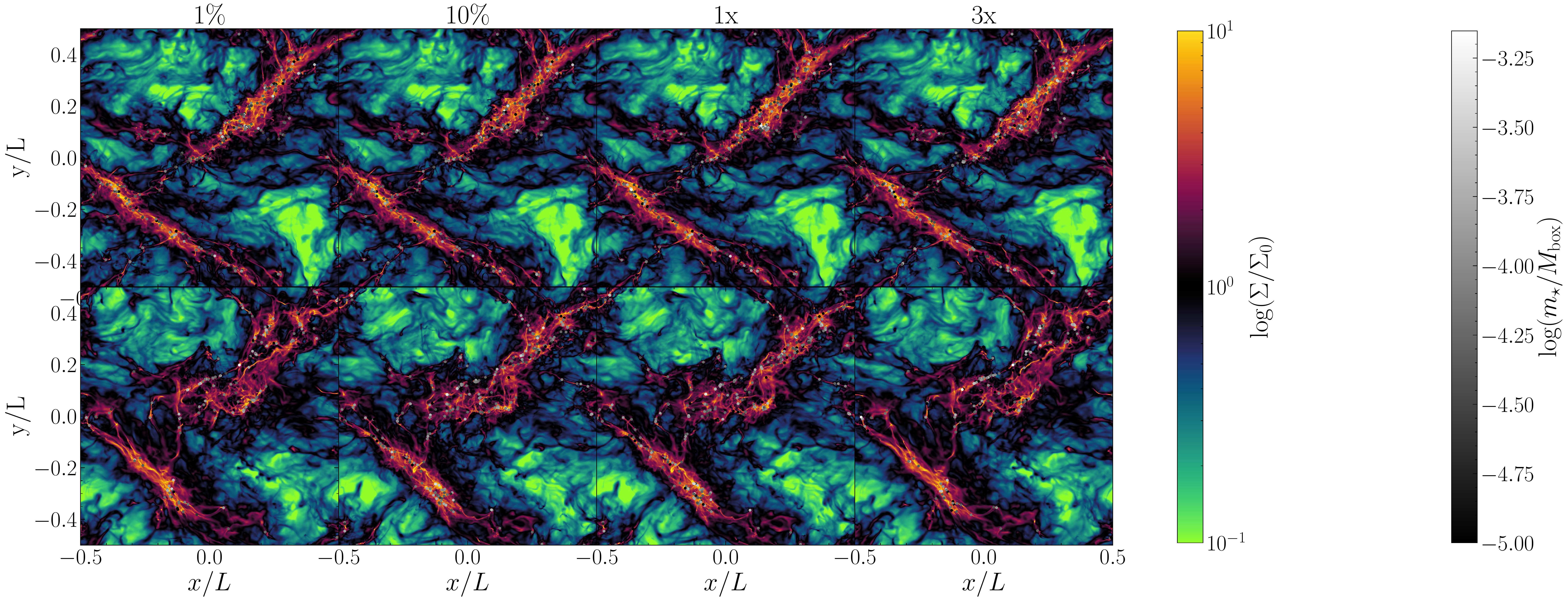}
    \caption{Column densities of simulations L (top) and M (bottom) series at 5\% SFE at metallicities 1\%, 10\%, 1$\times$, 3$\times$ solar metallicity. The colour scale goes from $\log(\Sigma/\Sigma_{0}) = -1$ to $1$, where $\Sigma_0 = \rho_{0} L$ and $\rho_0$ is the mean density in the simulation domain. Circles show star particles, and are colour-coded by mass $m_\star$ from $\log (m_{\star}/M_{\rm box}) = -5 $ to $-3$, where $M_{\rm box}$ is the total mass of the simulation box.} 
\label{fig:proj}
\end{figure*}

\begin{figure*}
	\includegraphics[width=\textwidth]{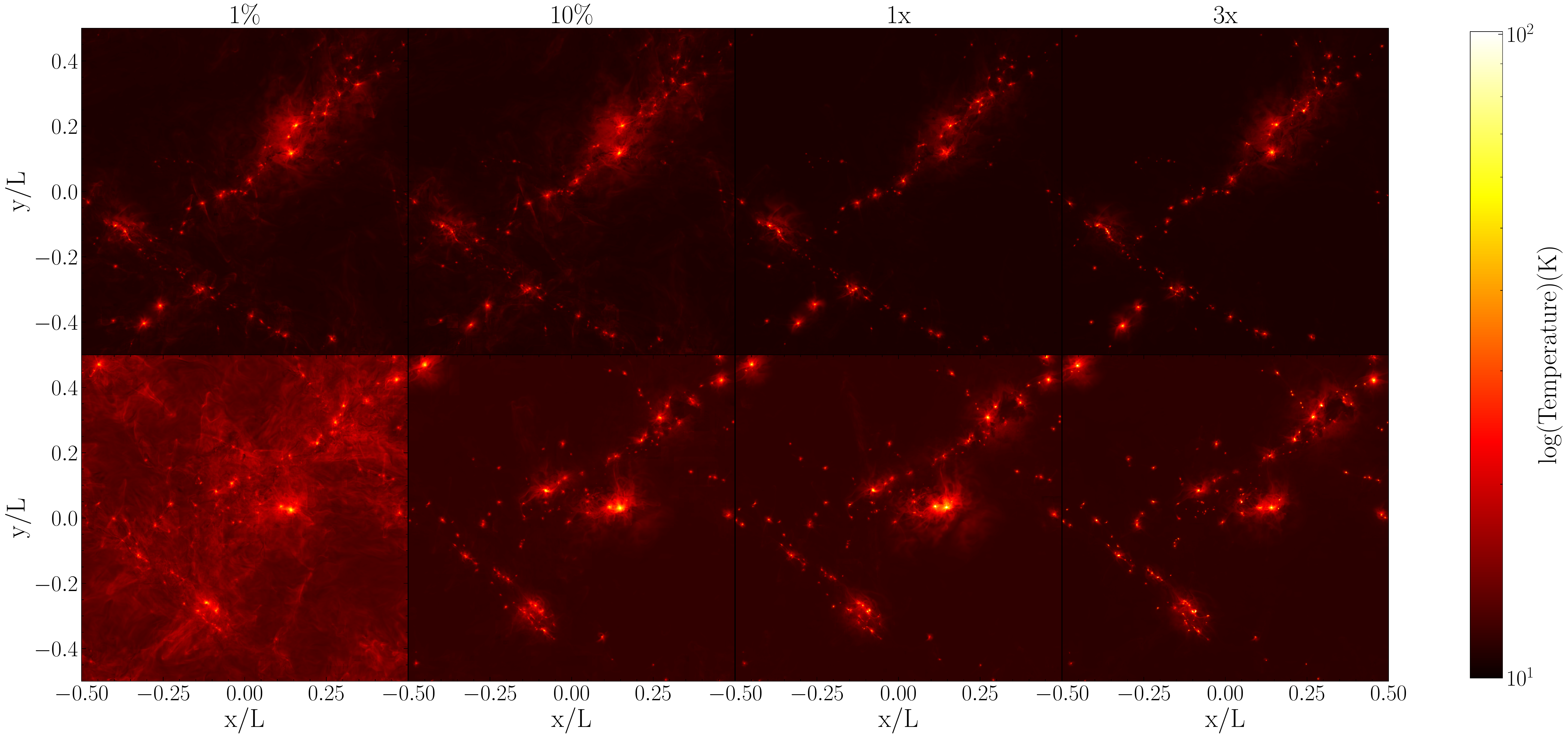}
    \caption{Same as \autoref{fig:proj}, but showing density-weighted projected temperature rather than column density.}
    \label{fig:projtemp}
\end{figure*}

In \autoref{fig:proj} and \autoref{fig:projtemp} we show the column density and the density-weighted temperature of runs L and M for our four different metallicities: 1$\%$, 10$\%$, $1\times$, and $3\times$ Solar metallicity. As in \citetalias{2022MNRAS.516.5712T} we show these plots at matching SFE (star formation efficiency) instead of at matching times since star formation occurs at different times in these runs. It is clear from \autoref{fig:proj} that turbulence has created dense filamentary structures and that star formation activity is confined within these structures. Morphologically the runs are very similar to one another, which is not surprising since by construction in the absence of a feedback mechanism these runs would be identical. There is one small difference visible between the two sets of runs with different column densities: run L produces a filamentary structure that is more straight and narrow than the filamentary structures produced in run M. By contrast, at different metallicities but fixed surface density the simulations are almost identical to one another. This indicates that, at least with regard to morphology, metallicity has a relatively minor impact.

The temperature structure of the gas shown in \autoref{fig:projtemp} behaves quite differently. Here differences with metallicity are much more apparent than in \autoref{fig:proj}. For both the L and M series, lower metallicity runs are warmer compared to the higher metallicity runs. This indicates that the lower metallicity runs stellar radiation is able to escape further from the dense regions around individual stellar sources, leading to more widely distributed heating and a warmer mean temperature over most of the volume. This is in contrast to the effects of surface density, in which lower surface density runs are cooler because radiation is trapped less efficiently in the full simulation box.

\begin{figure}
	\includegraphics[width=1.0\columnwidth]{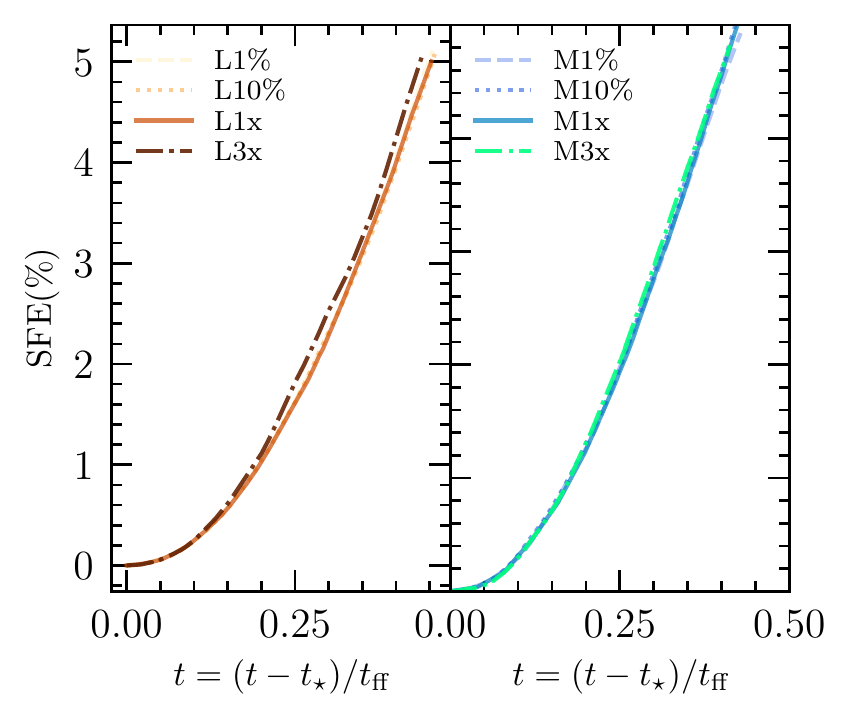}
    \caption{Star formation efficiency as a function of time since the formation of the first star $t_{\star}$, measured in units of the free-fall time $t_{\rm ff}$. The left panel shows the L series of runs at different metallicities, and the right panel shows the corresponding results for the M series.}
    \label{fig:sfetime}
\end{figure}

\begin{figure}
	\includegraphics[width=1.0\columnwidth]{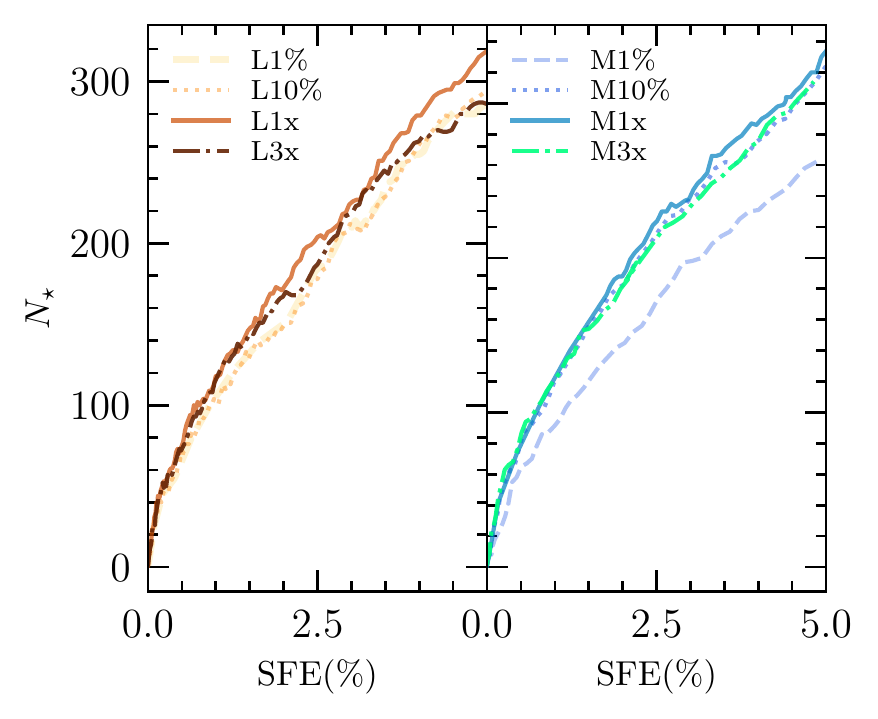}
    \caption{Number of stars formed as a function of star formation efficiency. As in \autoref{fig:sfetime}, the left panel shows the L run series and the right panel shows the M series.}
    \label{fig:nstarssfe}
\end{figure}

We show the time evolution of the star formation efficiency (SFE) in \autoref{fig:sfetime}, and the total number of stars present as a function of the SFE in \autoref{fig:nstarssfe}; we define the SFE as the ratio of total stellar mass in the simulation to the total initial mass present in the volume. We find that, in all runs, once the star formation activity begins it takes approximately 0.5 free fall times to reach 5$\%$ SFE. There is very little difference between the runs with varying metallicities at a fixed surface density. This indicates that whatever effects the feedback mechanisms have on the star formation rate in the simulations are independent of metallicity and surface density. By contrast, we see larger differences in \autoref{fig:nstarssfe}, which shows the number of stars as a function of SFE. At a fixed surface density, the two more metal-poor runs in both the L and M series, corresponding to 1\% and 10\% of Solar metallicity, produce the same number of stars at a given SFE. On the other hand, the runs with Solar and $3\times$ Solar metallicity produce more stars at fixed SFE, indicating that the stars formed in these runs have systematically lower masses; the effect is not large, but is clear, particularly in the L runs. We examine these trends further in the next section.

\subsection{Stellar mass distribution}
\label{ssec:massdist}
 \begin{figure}
	\includegraphics[width=1.0\columnwidth]{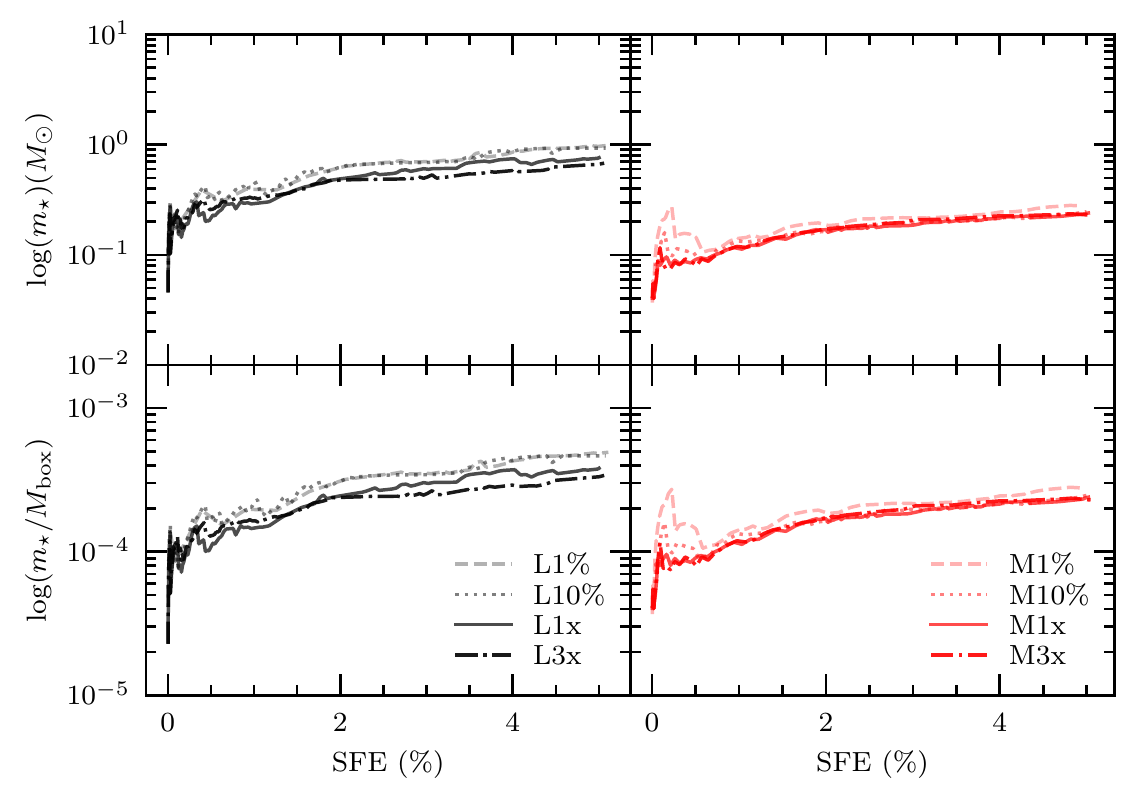}
    \caption{The top panels shows the evolution of the median of the sink particle mass distribution for run L (left) and M (right) at different metallicities in absolute mass expressed in M$_\odot$. The lower panel shows the evolution relative to the box mass $M_{\rm box}$. Here the medians are calculated with respect to mass rather than number.}
\label{fig:median}
\end{figure}

To explore the metallicity dependence of the stellar mass function further, we present two sets of plots. The first, \autoref{fig:median}, shows the evolution of the median of the sink particle mass distributions for all the runs. As in \citetalias{2022MNRAS.516.5712T} we measure the median with respect to the stellar mass rather than the number, i.e., the median stellar mass $m_*$ is defined by the condition that half the total stellar mass is found in stars with masses $<m_*$. Looking at \autoref{fig:median} we see that the median masses have reached nearly steady values at 5$\%$ SFE. When measured on both absolute (top panel) and relative scales (bottom panel) the L runs at varying metallicity show greater variation in median mass than the M runs; however, the direction of variation is the same in both sets of runs, which is that as the metallicity increases the median mass decreases. However, we further note that the differences between the different metallicities are smaller than the differences between runs L and M, particularly when expressed in terms of absolute rather than relative mass. 

\begin{figure}
	\includegraphics[width=1.0\columnwidth]{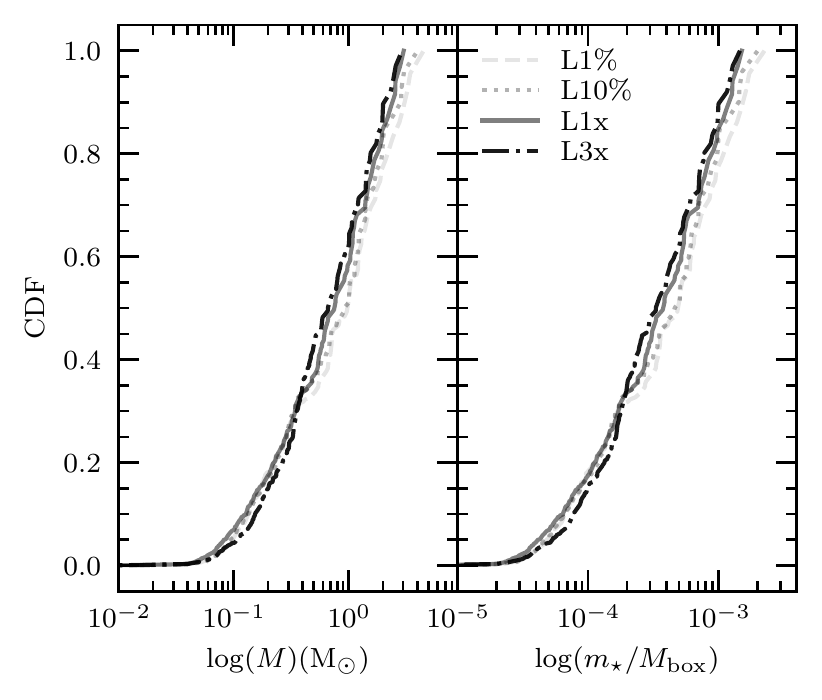}
	\includegraphics[width=1.0\columnwidth]{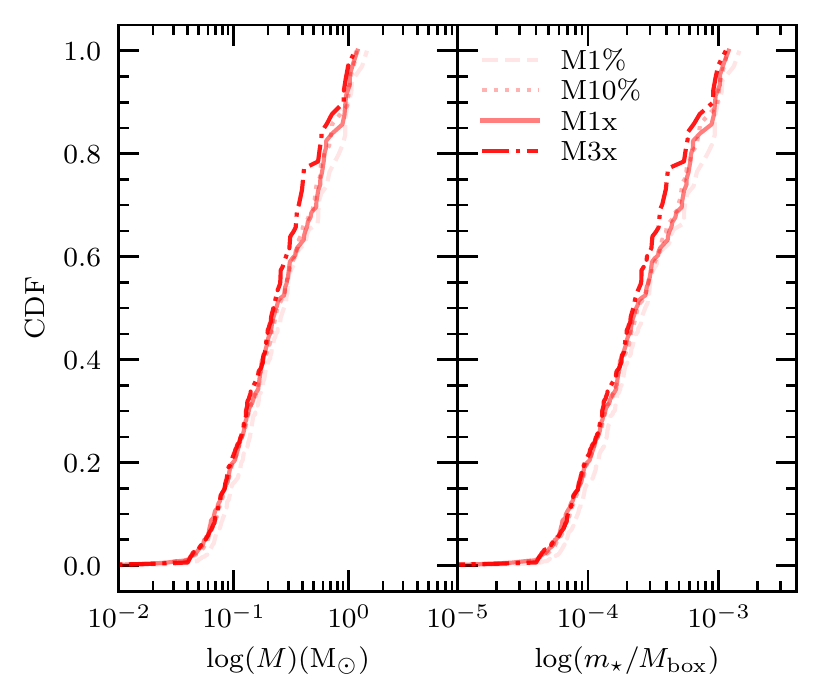}
    \caption{Cumulative distribution function (CDF) of the sink particle masses of the simulations, the top panel is the run L at different metallicities, and the bottom panel is run M at different metallicities. The left side is the CDF with respect to absolute stellar mass and the right side is the CDF for mass measured relative to the box mass $M_{\rm box}$. In both the L and M series, lower metallicity runs produce a slightly top-heavy IMF than the higher metallicity runs}
\label{fig:cdf_plot}
\end{figure}

To investigate these differences further, in \autoref{fig:cdf_plot} we show the cumulative mass functions of the runs at 5$\%$ SFE on both absolute and relative mass scales; as in \autoref{fig:median}, we measure the CDF with respect to mass rather than number, since this is much more numerically stable. Consistent with the trend as a function of time shown in \autoref{fig:median}, in the L runs as the metallicity increases there is a slight shift in the CDF shape and mass range it covers, with metal-poor runs showing slightly heavier mass distributions than metal-rich runs at the high mass end. The trend is less visible, and may in fact reverse, at the lowest stellar masses, so that the metal-rich distribution is narrowest. The variations in run M are similar to or perhaps slightly smaller than those in run L, and are in qualitatively the same direction. Consistent with \citetalias{2022MNRAS.516.5712T}, we also see that, on an absolute scale, the L runs have a slightly heavier IMF than the M runs. 

\subsection{The effect of radiation feedback at varying metallicity}
\label{ssec:radiation}

Recall that our simulation series is constructed so that if the gas were isothermal and not subject to protostellar feedback, the different metallicity and surface density runs would all simply be re-scaled versions of the same simulation, and we would therefore expect identical results. Thus the differences in IMF we have observed between the runs must be a result of feedback, and variations in IMF at fixed surface density must be due to the interaction of feedback with the gas of varying metallicity. In this section, we examine how radiation feedback alters gas temperature distribution in runs of varying metallicity and how this, in turn, influences fragmentation, and in the next section, we perform a similar exercise for protostellar outflow feedback.

\begin{figure*}
	\includegraphics[width=\textwidth]{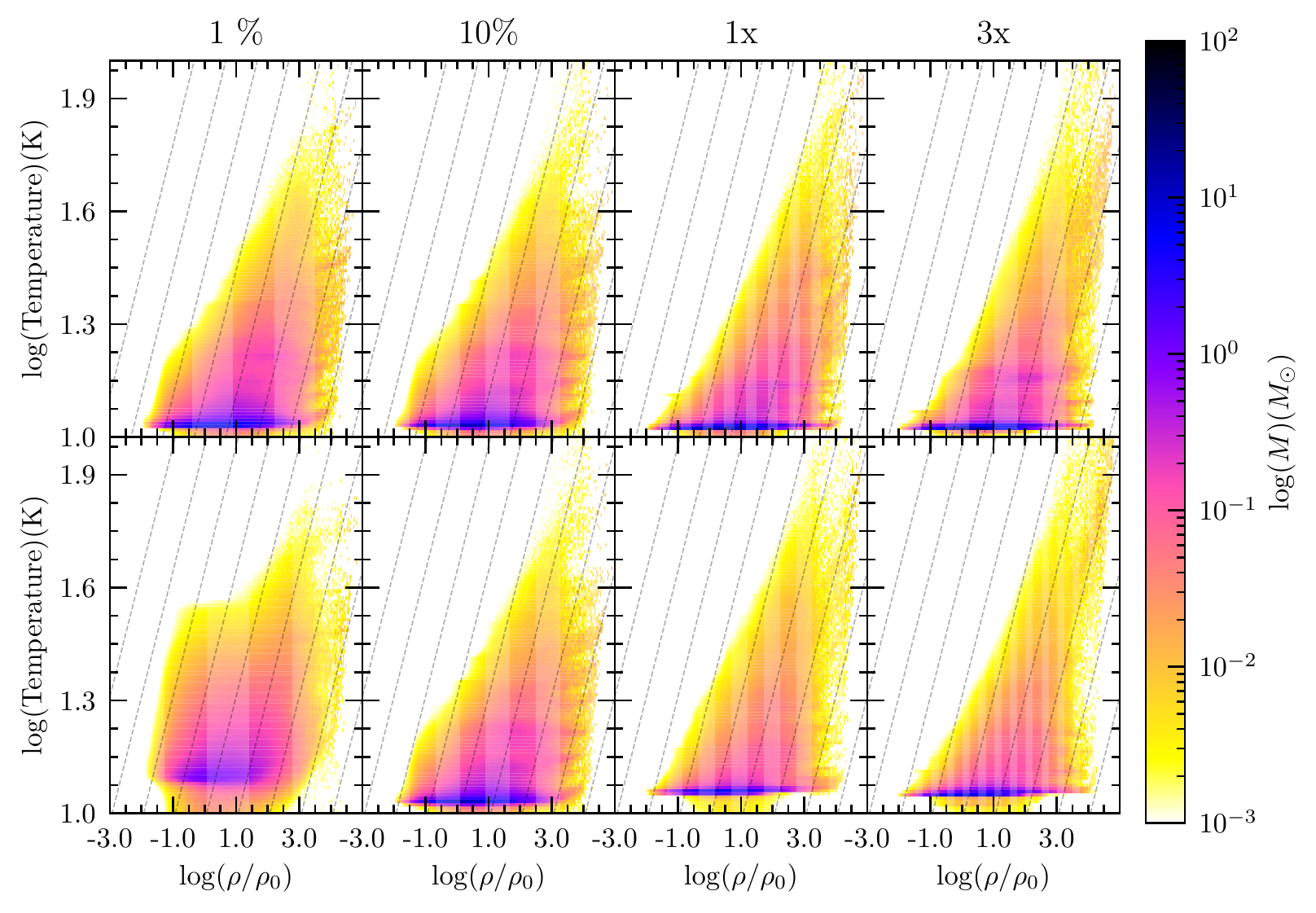}
	
    \caption{The joint distribution of normalised density $\rho/\rho_0$ and temperature $T$ for runs L and M at different metallicity. The top row shows the state of the L runs and the bottom row shows the M runs; metallicity varies from 1\% to 3$\times$ Solar from left to right, as indicated above the columns. All plots show the state of the simulations when they reach 5\% SFE. The colour bar shows the mass in each density-temperature bin. Dashed lines indicate loci of constant box-normalised Jeans mass $M_{\rm J}/M_{\rm box}$; lines are spaced logarithmically at intervals of 0.5 dex, with the left-most line corresponding to $\log M_{\rm J}/M_{\rm box} = -2$.}
    \label{fig:phase2d}
\end{figure*}

\autoref{fig:phase2d} shows the gas mass distribution with respect to both density and temperature in runs L and M at different metallicity, all at 5\% SFE. From the plot, it is clear that there are differences between the metal-poor and the metal-rich runs. The most significant differences are in the high-density parts of the distribution since this is the gas that is closest to star formation and thus whose fragmentation will most directly affect the IMF. In the metal-poor runs, we see an upturn in the minimum temperature of the dense gas, while in the more metal-rich runs the temperature remains flat as high density. This difference occurs because at low metallicity the gas is less opaque and therefore is not able to shield itself from heating by the radiation produced by nearby stars, and therefore becomes warmer as it collapses. 
The most extreme case here is the M run at 1\% of Solar metallicity, where there is essentially no gas at densities $>10^{3} \rho_{0}$ and temperatures $\lesssim 15$ K. On the other hand, the temperature distribution at high density extends to noticeably higher temperatures in the metal-rich runs than in the metal-poor ones, most likely for the same reason: because when the gas is more opaque, radiation is more easily ``bottled up'', leading to hotter warm regions around protostars that have already started radiating. While all of these temperature differences might seem minor,  recall that the Jeans mass varies as $T^{3/2}$, so a factor of $1.5$ corresponds to a factor of $1.8$ in mass -- comparable to or larger than the IMF shifts we have observed between the runs, and of roughly the  size required to explain the observations of ETGs.

\begin{figure}
	\includegraphics[width=1.0\columnwidth]{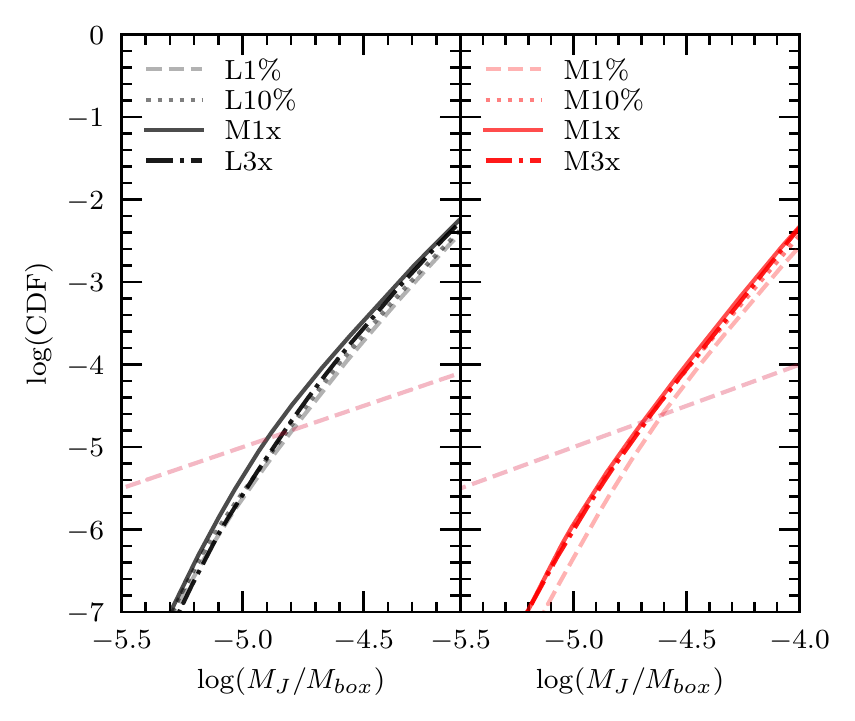}
	
    \caption{Cumulative distribution functions of mass with respect to $M_{\rm J}/M_{\rm box}$ for the L  run series (left) and the M run series (right). The dashed straight lines in each panel show the one-to-one relation, and indicate a minimum condition for fragmentation: for any mass $M_{\rm J}/M_{\rm box}$ for which the CDF falls below the dashed line, there is less than a single Jeans mass of material with $M_\mathrm{J}$ that small in the box, and thus it is impossible to create an object of that mass via gravitational collapse.}
    \label{fig:jeansCDF}
\end{figure}

In order to understand how the opacity affects fragmentation more directly, we show in \autoref{fig:phase2d} the 1D mass-weighted cumulative distribution function of mass with respect to $M_{\rm J}/M_{\rm box}$, where $M_\mathrm{J}$ is the Jeans mass computed from the local gas density and temperature. For a detailed description of how to construct this figure we refer readers to \citetalias{2022MNRAS.516.5712T}), but to summarise, note that the horizontal axis in \autoref{fig:jeansCDF} corresponds to the dashed lines of constant $M_\mathrm{J}/M_\mathrm{box}$ in \autoref{fig:phase2d}. The vertical axis in \autoref{fig:jeansCDF} then shows what fraction of the mass in the simulation lies to the right of the corresponding dashed line in \autoref{fig:phase2d}, i.e., it shows that fraction of the mass in a given simulation has a Jeans mass smaller than the indicated value, and thus has the potential to fragment to produce stars with mass smaller than that value. From this plot, we see a clear trend between the metal-poor and metal-rich runs. The metal-rich runs tend to have a more mass at lower $M_{\rm J}/M_{\rm box}$ than the metal-poor runs, consistent with our observation that the metal-rich runs to produce a more bottom-heavy IMF. The location at which the dashed 1-1 line in the figure intersects the CDF gives a rough estimate of the minimum object mass that is possible for gravity to produce: to the left of this intersection point, the box contains less than a single Jeans mass of material for which $M_J$ is that small, and is therefore not capable of producing collapsed objects of such small mass. The intersection point is on average shifted to slightly lower mass at higher metallicity, indicating that radiation feedback is more efficient in metal-poor runs at suppressing the formation of lower-mass objects.

\subsection{The effect of outflow feedback}
\label{ssec:outflow}

\begin{figure}
	\includegraphics[width=1.0\columnwidth]{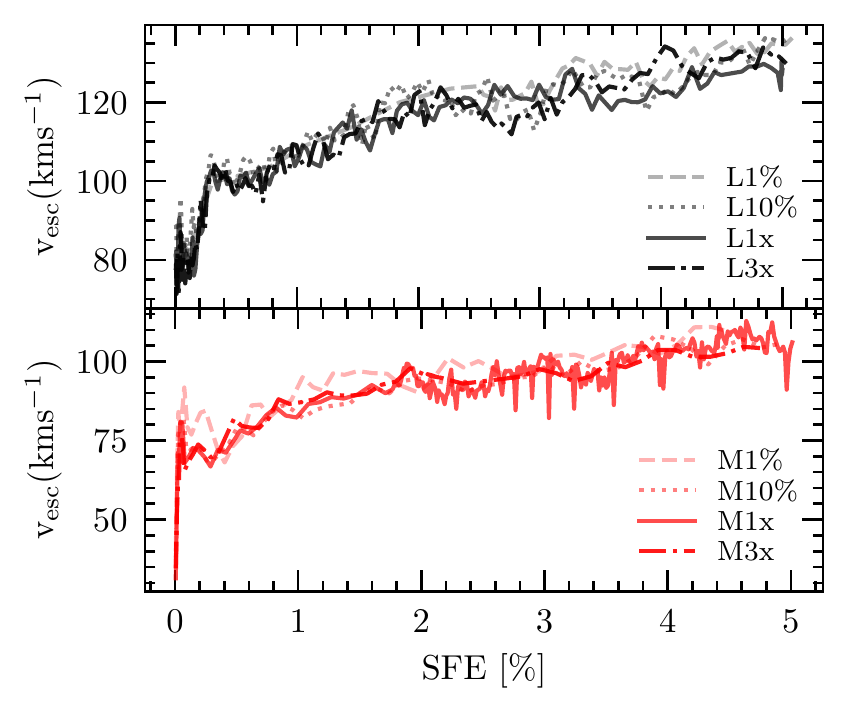}
	\caption{Accretion rate-weighted mean surface escape speed of stars formed the simulations as a function of SFE. The top panel shows the L series of run at varying metallicity, and the bottom panel shows the M series.}
    \label{fig:vesc}
\end{figure}

\begin{figure}
	\includegraphics[width=1.0\columnwidth]{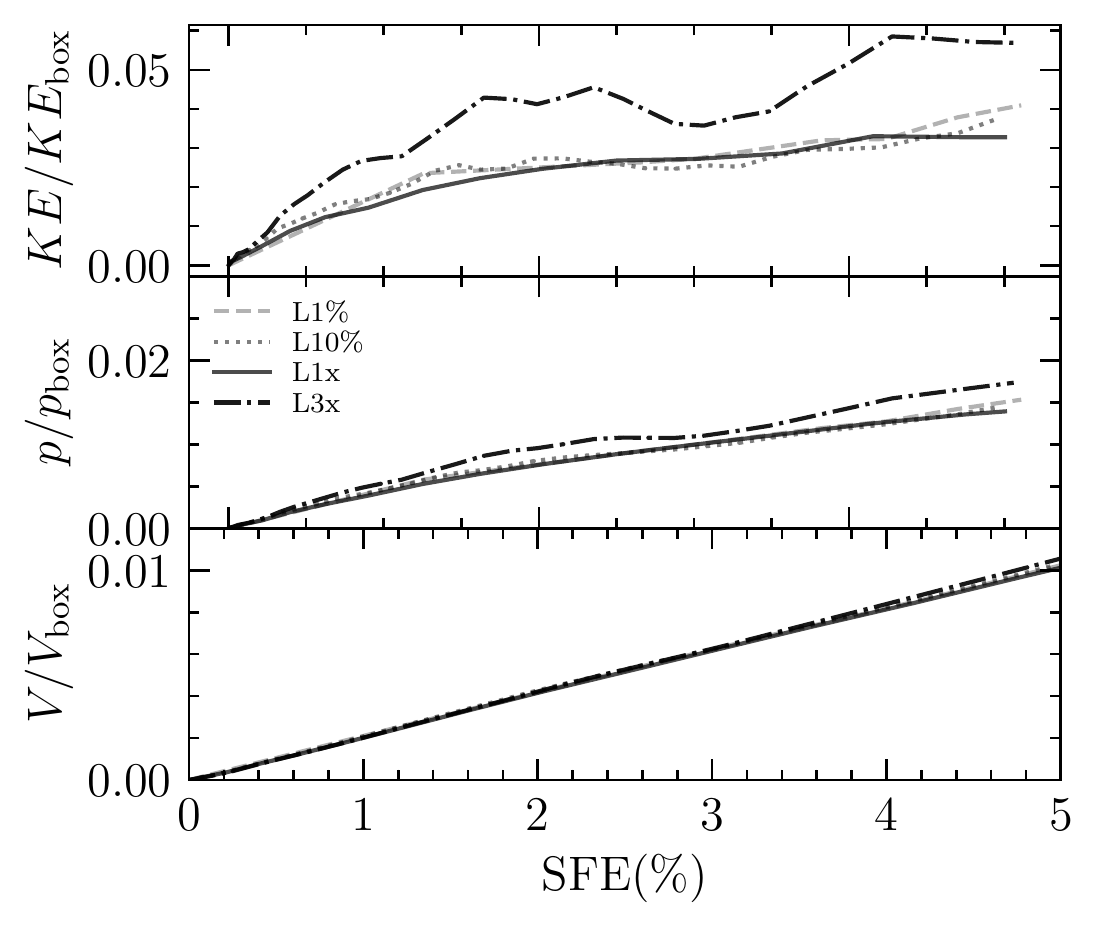}
 \includegraphics[width=1.0\columnwidth]{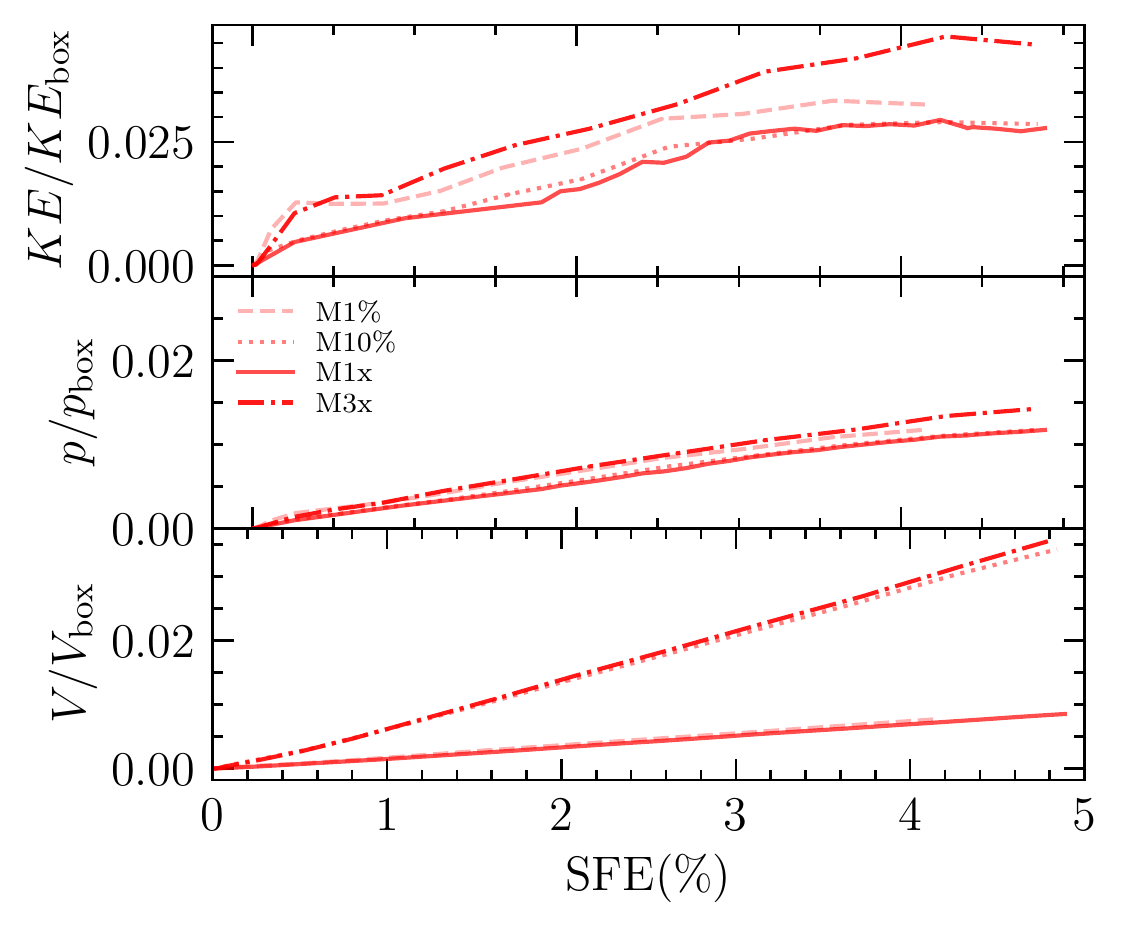}
	\caption{Outflow kinetic energy, momentum, and volume relative to the box kinetic energy, momentum, and volume, all as a function of SFE. The top set of panels shows the L series of runs and the bottom shows the M series. }
    \label{fig:outflow}
\end{figure}

As discussed in \citetalias{2022MNRAS.516.5712T}, radiation and outflow feedback play complementary roles in shaping the IMF: the former sets a lower bound on the masses of stars that can form, shaping the low-mass end of the IMF, while the latter inhibits the accumulation of mass by the most massive stars, shaping the high-mass end of the IMF. This difference explains why it is possible for the low-mass and high-mass parts of the CDF to shift in opposite directions, as we observe in \autoref{fig:cdf_plot}. In this case, if radiation feedback were the only actor in our simulations, then we would expect a uniform shift of the IMF towards the higher masses in the metal-poor runs. The actual pattern of IMF shift with metallicity is more complex, indicating that protostellar outflow feedback is significant as well, and that its dependence on metallicity is not identical to that of radiation feedback. 

To understand how outflow feedback affects our runs we first look at \autoref{fig:vesc}, which shows the evolution of accretion rate-weighted mean surface escape speed in the runs as a function of SFE. This quantity matters because the outflow prescriptions we use in these simulations link the outflow velocity to the stellar surface escape speed, as is observed to be the case \citep[and references therein]{2000prpl.conf..867R} and as is expected theoretically \citep[and references therein]{Konigl00a}, since outflows are launched from close to the stellar surface. See \citet{2011ApJ...740..107C} for a detailed description of our outflow prescription. From the plot, it is clear that $v_\mathrm{esc}$ varies with surface density (run L has a higher surface escape speed than run M) but for a fixed surface density there is no systematic difference in the surface escape speed between the different metallicity runs; this is not surprising, given that the primary determinant of escape speed is the ratio of the time stars have had to contract toward the main sequence to the stellar Kelvin-Helmholtz timescale; the former depends on surface density but not on metallicity (since the simulation star formation rates are almost independent of metallicity), and the latter to good approximation depends on neither. Thus we find that at any given metallicity the L series of runs has higher mean escape speed than the M series, since the L runs take longer to reach a given SFE, but that there is no difference with metallicity.

Thus the differences we find in how outflows influence the IMF at different metallicity cannot be a result of differences in the outflow strength. Instead, in order to understand how protostellar outflows are breaking the symmetry in the simulations, we next examine \autoref{fig:outflow}, which shows the total kinetic energy, scalar momentum, and volume occupied by outflow material normalised to the box kinetic energy ($E_\mathrm{box} = M_{\rm box} \sigma_v^2/2$, where $\sigma_v$ is the initial 3D velocity dispersion), scalar momentum by $p_\mathrm{box} = M_{\rm box}\sigma_v$ and volume $V_\mathrm{box} = L^3$; we identify outflow material by a passive scalar that we add to the gas launched into outflows, and refer readers to \citetalias{2022MNRAS.516.5712T} for details on the procedure. Looking at this plot we see that in the metal-rich runs the outflowing material on average has more kinetic energy and scalar momentum, and occupies more volume, than for the more metal-poor runs at the same surface density. In \citetalias{2022MNRAS.516.5712T} we found that greater outflow momentum results in \textit{less} efficient fragmentation of the gas, because the outflows punch through their environments and escape more efficiently, therefore creating an environment more favourable to the formation of higher-mass stars. However, this does not appear to be the case here: the metal-rich runs have more outflow momentum and kinetic energy, but have slightly \textit{fewer} massive stars, as can be seen by examining the CDFs shown in \autoref{fig:cdf_plot}. One possible explanation is that the outflow momentum and kinetic energy vary between the L and M run series for different reasons than they vary with metallicity at fixed surface density. The outflow energy and momentum are larger in the L runs than the M ones because the L runs have more powerful outflows due to the higher surface escape speeds achieved by stars that have longer to contract toward the main sequence. By contrast, the differences between the runs at different metallicities are almost certainly driven by the different cooling rates of the outflow gas and the dense cloud material with which it mixes. In the metal-poor runs, cooling times are longer due to the lower metal content of the gas, and as a result, regions of shocked cloud gas may build up that pressure-confine the outflow more effectively, in turn making outflows less efficient at breaking up the gas clumps. This allows slightly more massive stars to form at the high-mass end of the IMF, as we observe in \autoref{fig:cdf_plot}. 


\subsection{Implications for the mass to light ratio in early type galaxies}
\label{ssec:mass_to_light}

One of the primary motivations of this study is to investigate the role of metallicity in the variations of the IMF that have been observed in ETGs. The mass-to-light ratio is the most direct line of evidence we can collect from our simulations to study these variations. In this section we explore the mass-to-light ratios of our simulations with varying metallicity. We use the \textsc{slug} stellar population synthesis code \citep{da-Silva12a, Krumholz15b} to generate isochrones at stellar population ages of 5 to 10 Gyr and at the metallicities used in the simulations (see \citetalias{2022MNRAS.516.5712T} for a detailed description of the procedure). We use the isochrones to calculate the mass-to-light ratio of the stellar populations formed in our simulations at ages of 5 to 10 Gyr in the SDSS $r$ band, which is commonly used for measurements of $M/L$ in ETGs.

\begin{figure}
	\includegraphics[width=1.0\columnwidth]{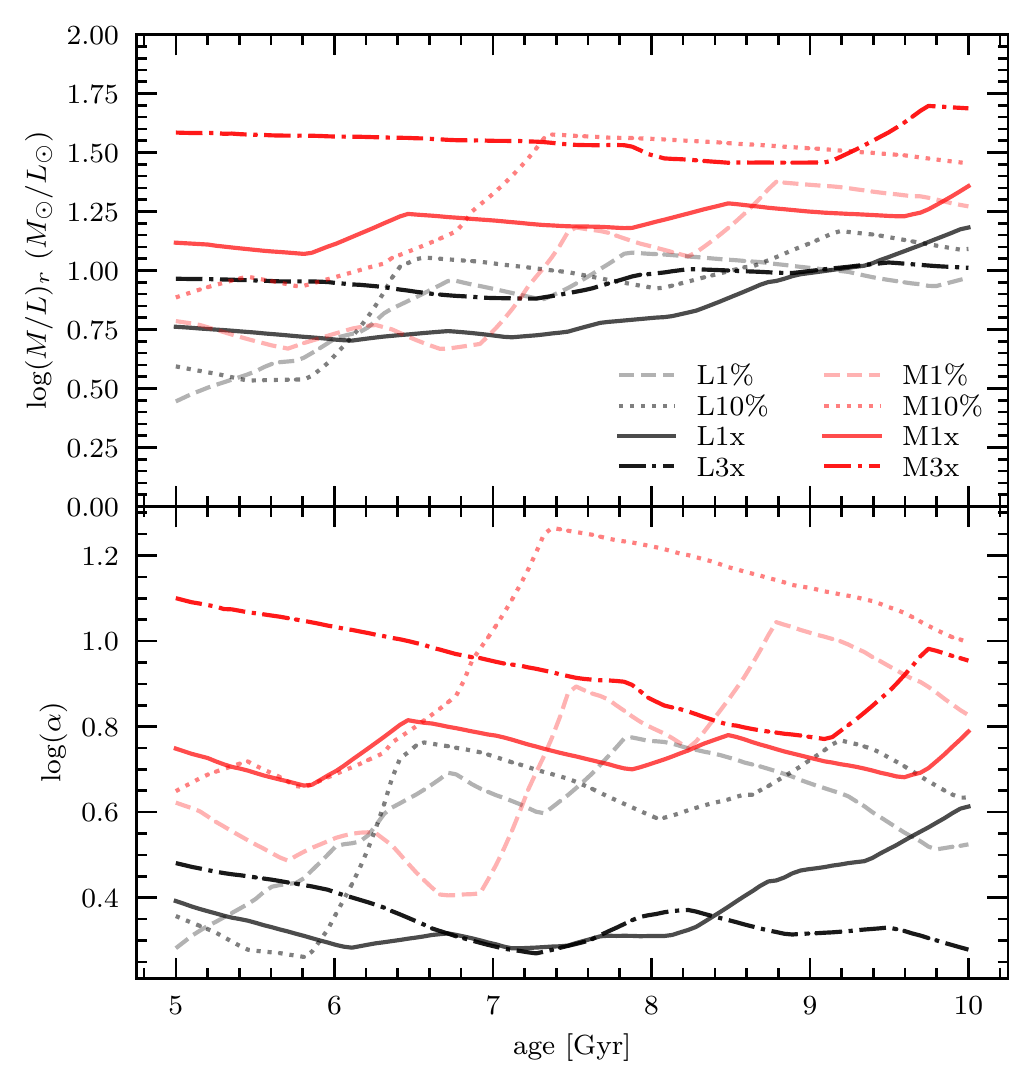}
    \caption{The top panel shows the mass to light ratio $(M/L)_r$ in the SDSS $r$ band as a function of population age computed for the stellar population in formed in each of the simulations, using isochrones computed with the same metallicity as used in the simulations. The black lines represent the $L$ series of runs and the red lines represent the $M$ series runs, respectively, with different line styles corresponding to different metallicity as indicated in the legend. The bottom panel shows the IMF mismatch parameter $\alpha = (M/L)/(M/L)_{Z_\odot,\rm Chabrier}$, where $(M/L)_{Z_\odot,\mathrm{Chabrier}}$ is the mass to light ratio expected for a Solar-metallicity stellar population with a \citet{2005ASSL..327...41C} IMF, again as a function of stellar population age.}
    \label{fig:ml_plot}
\end{figure}

The top panel of \autoref{fig:ml_plot} shows the mass-to-light evolution of the runs from age 5 to 10 Gyr, while the lower panel shows the IMF mismatch parameter $\alpha$, defined as the ratio of the actual $M/L$ ratio in the simulations to the $M/L$ ratio expected for a population with a Solar metallicity and a \citet{2005ASSL..327...41C} IMF at the same age; this latter quantity is the index most commonly used in observations to study IMF variations in ETGs, though we emphasise that the absolute value here is less significant than the differences between the runs due to the systematic uncertainties. For both the absolute $M/L$ or the IMF mismatch parameter, we see that, at any given metallicity, there is a difference of $\sim 0.3-0.5$ dex in mass to light ratio between the L and M runs; this finding is consistent with that in \citetalias{2022MNRAS.516.5712T} for the Solar metallicity case, and extends the results to non-Solar metallicities. At a fixed surface density, by contrast, there is a smaller difference in the mass-to-light ratio between the runs, consistent with our finding that surface density is a more influential factor than metallicity when it comes to determining the IMF.

\begin{figure}
	\includegraphics[width=1.0\columnwidth]{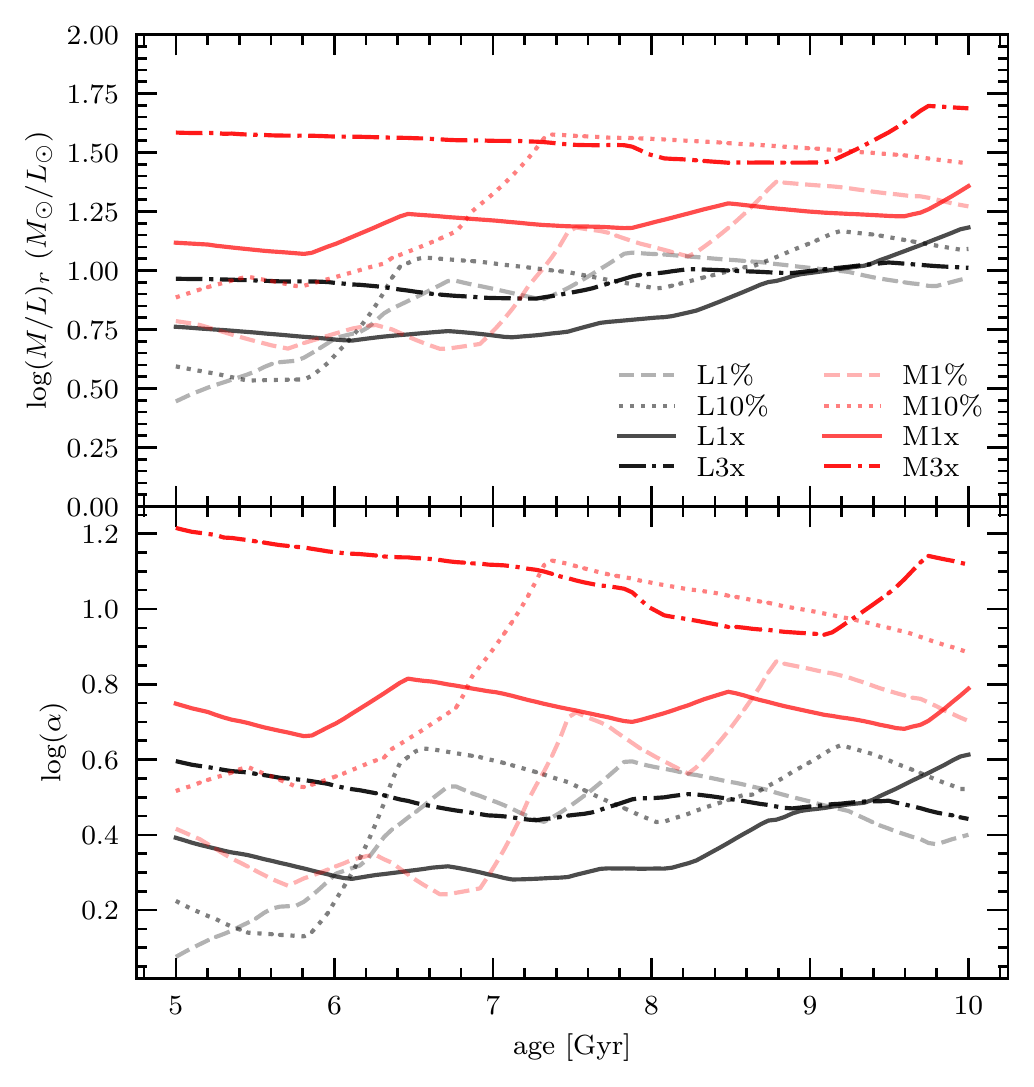}
    \caption{Same as \autoref{fig:ml_plot}, but now computed using the same set of isochrones (corresponding to those for $Z=Z_\odot$ for all simulations), regardless of the simulation metallicity.}.
    \label{fig:ml_plot1}
\end{figure}

Moreover, \autoref{fig:ml_plot} in fact somewhat exaggerates the effect of metallicity, because some of the difference in mass-to-light ratio between the differing metallicity runs occurs because the metallicity itself affects stellar evolution and atmospheres; thus we would not expect to identical $M/L$ values for two populations of different metallicity even if they had exactly the same IMF. To remove this effect, in \autoref{fig:ml_plot1} we show the same quantities as in \autoref{fig:ml_plot}, but where we have calculated the mass-to-light ratio for all runs using the isochrones for Solar metallicity; while this is clearly artificial, it enables us to isolate the effects of metallicity on the IMF from the effects on stellar evolution and atmospheres. In \autoref{fig:ml_plot1} the runs at different metallicity but the same surface density cluster even more tightly, and are further from the corresponding runs at different surface density, than in \autoref{fig:ml_plot}. This reinforces our conclusion that the effects of metallicity on the IMF and the observational diagnostics used to assess it are fairly minor compared to the effects of surface density.

Our qualitative conclusion that metallicity effects cannot be responsible for the IMF variations seen in ETGs is the same as that reached in a recent paper by \cite{2023MNRAS.519..688B}, who explored the role of varying metallicity on the stellar properties at redshift $z =5$. \citeauthor{2023MNRAS.519..688B} found that at this redshift, with its higher CMB temperature, increasing metallicity leads to an increase in the characteristic mass of stars, exactly the opposite of what would be required to reproduce observed IMF variations in ETGs. However, our conclusions differ somewhat in detail; \citeauthor{2023MNRAS.519..688B} varied both the metallicity and the CMB temperature, whereas we, in keeping with our philosophy of controlled experiments, have varied only the former, a decision that is likely to be significant at $z\gtrsim 3$, when the CMB temperature begins to exceed our adopted background IR radiation field temperature of 10 K. Thus there is no contradiction between the findings of \citet{2023MNRAS.519..688B} and of this work, since the experiments we have carried out are different. On the other hand, our finding that there is a very weak IMF variation with metallicity at fixed surface density is qualitatively consistent with the findings of \citet{Bate19a}, though they explore only a single surface density case, one similar to our L series.

\section{Conclusions}
\label{Conclusion}

In this paper, we present a set of radiation-magnetohydrodynamic simulations of star formation including radiation and protostellar outflow feedback from young stars. We carry out a systematic exploration of how the mass function of the stars formed in the simulations varies as a function of surface density and metallicity, exploring surface densities from those typical of Milky Way-like galaxies to those typical of starbursts or early-type galaxies, and metallicities that range from those typical of the most metal-poor local dwarfs, $Z = \mathrm{Z}_\odot/100$, to those typical of the centres of the most metal-rich early-types, $Z= 3\mathrm{Z}_\odot$. The setup of the simulations allows us to separate out the effects of metallicity at fixed surface density and of surface density at fixed metallicity, and to identify the specific mechanisms by which these parameters interact with stellar feedback. This extends the results of \citet{2022MNRAS.516.5712T}, who used the same methodology to explore the effects of surface density alone, without metallicity variation. As in that paper, they key to approach is to set up a series of simulations where we keep all dimensionless parameters except the cloud optical depth (which depends on both surface density and metallicity) constant, so that, for isothermal gas in the absence of stellar feedback, all the simulations would just be rescaled versions of one another, and would produce identical results. 

Overall we see a trend whereby metal-poor cases produce a slightly heavier mass distribution over most of the IMF than metal-rich cases. However, this trend disappears or reverses at the lower mass end of the IMF, so that metal-rich cases also have a slightly narrower IMF overall. We attribute these shifts to the contrasting ways that metallicity variation interacts with different types of stellar feedback. This high mass end of the IMF is sensitive primarily to the effects of protostellar outflow feedback inhibiting the most massive objects from accreting, something that appears to happen slightly more efficiently when the metallicity is higher and cooling of outflow-shocked gas is more rapid, suppressing the formation of more massive objects. By contrast the remainder of the IMF is shaped primarily by radiation feedback suppressing fragmentation and preventing small objects from forming, something that appears to happen slightly more efficiently at lower metallicity. The net effect is that the IMF shifts to somewhat lower masses, and is also somewhat narrower, when the metallicity is higher.

While these differences with metallicity provide interesting insight into how feedback interacts with the star-forming environment, they are ultimately of rather minor importance. This is because on an absolute scale the biggest differences we see in the IMF are set by variations in surface density. Our low surface density cases produce a slightly heavier IMF than our higher surface ones, and the differences are substantially larger than the subtle variations induced by metallicity. When we explore how these IMF variations compare to those required to explain observed variations in the mass-to-light ratio in early-type galaxies compared to spiral galaxies, we find that metallicity-induced IMF changes are too small to explain the observations, whereas surface density-induced ones are at the right level. We, therefore, conclude that surface density is a more important factor than metallicity in determining the stellar IMF and that the differences observed between the IMFs of spiral and early-type galaxies are most likely an effect of interstellar pressure and therefore surface density, not an effect of metallicity.

\section*{Acknowledgements}

TST and MRK acknowledge support from the Australian Research Council through Laureate Fellowship award FL220100020. This research was undertaken with the assistance of resources and services from the National Computational Infrastructure (NCI), which is supported by the Australian Government, and by resources provided by the Pawsey Supercomputing Research Centre with funding from the Australian Government and the Government of Western Australia.

\section*{Data Availability}

The data underlying this article will be shared on reasonable request to the corresponding author.



\bibliographystyle{mnras}
\bibliography{mnras_template} 





\bsp	
\label{lastpage}
\end{document}